\documentclass[12pt]{article}

\usepackage{amsmath}
\usepackage{amssymb}
\usepackage{graphicx, subfig}
\usepackage{url}
\usepackage{multirow}
\usepackage{authblk}
\usepackage{mathtools,blkarray}
\usepackage{hyperref}
\usepackage[margin=1.25in]{geometry}
\usepackage{natbib}
\usepackage{adjustbox}

\usepackage{tikz}
\usetikzlibrary{shapes.geometric, arrows}
\usetikzlibrary{shapes.geometric, arrows}
\tikzstyle{startstop1} = [rectangle, rounded corners, minimum width=0.5cm, minimum height=0.6cm, text centered,text width=3cm, draw=black, fill=red!30]
\tikzstyle{startstop2} = [rectangle, rounded corners, minimum width=0.3cm, minimum height=0.6cm, text centered,text width=3cm, draw=black]
\tikzstyle{startstop} = [minimum width=0.5cm, minimum height=0.5cm, text centered, draw=black,fill=red!30]
\tikzstyle{inter} = [minimum width=1.3cm, minimum height=0.8cm, text centered,draw=black]
\tikzstyle{inter1} = [rectangle, rounded corners, minimum width=3cm, minimum height=1.5cm, text centered,text width=4cm, dashed, draw=black]
\tikzstyle{arrow} = [thick,->,>=stealth]

\begin{document}

\title{Instrumental variable estimation of early treatment effect in randomized screening trials}

\author[1]{Sudipta Saha}
\author[2]{Zhihui (Amy) Liu}
\author[1]{Olli~Saarela\thanks{Correspondence to: Olli Saarela, Dalla Lana School of Public Health, 155 College Street, Toronto, Ontario M5T 3M7, Canada. Email: \texttt{olli.saarela@utoronto.ca}}}

\affil[1]{Dalla Lana School of Public Health, University of Toronto}
\affil[2]{Princess Margaret Cancer Centre, University Health Network}

\maketitle

\begin{abstract}

The primary analysis of randomized screening trials for cancer typically adheres to the intention-to-screen principle, measuring cancer-specific mortality reductions between screening and control arms. These mortality reductions result from a combination of the screening regimen, screening technology and the effect of the early, screening-induced, treatment. This motivates addressing these different aspects separately. Here we are interested in the causal effect of early versus delayed treatments on cancer mortality among the screening-detectable subgroup, which under certain assumptions is estimable from conventional randomized screening trial using instrumental variable type methods. To define the causal effect of interest, we formulate a simplified structural multi-state model for screening trials, based on a hypothetical intervention trial where screening detected individuals would be randomized into early versus delayed treatments. The cancer-specific mortality reductions after screening detection are quantified by a cause-specific hazard ratio. For this, we propose two estimators, based on an estimating equation and a likelihood expression. The methods extend existing instrumental variable methods for time-to-event and competing risks outcomes to time-dependent intermediate variables. Using the multi-state model as the basis of a data generating mechanism, we investigate the performance of the new estimators through simulation studies. In addition, we illustrate the proposed method in the context of CT screening for lung cancer using the US National Lung Screening Trial (NLST) data.

\end{abstract}

\section{Introduction}
The benefits of cancer screening are ideally studied through randomized screening trials, where asymptomatic participants are assigned to either undergo a prespecified regimen of screening examinations, or to control without screening, and then followed up for cancer specific mortality \citep{Hu1997}. The primary analysis of such a trial usually follows the intention to screen (ITS) principle, based on comparison of cancer specific mortality between the screening and control arms at the end of the follow-up period. Any mortality reductions in the screening arm are due to the combination of earlier (asymptomatic) detections due to the screening technology, regimen and uptake, and the effectiveness of the subsequent early (versus delayed, following symptomatic diagnosis) treatments. This motivates studying these two aspects separately. The aim in isolating the effect of the early treatment is to obtain an effect measure that is less dependent on the screening regimen and uptake and potentially more transferable between populations with for example different disease incidence. Herein we are interested in the effectiveness of the early treatments in the subpopulation who would be screening detectable if offered screening. The subgroup effectiveness can be studied based on data collected through conventional screening trials, using instrumental variable type estimators, where the instrument is the randomized assignment. 

Instrumental variable (IV) approach has been primarily introduced to adjust for non-compliance with the treatment assignments in therapeutic trials \citep{baker1994paired,imbens1994identification,angrist1996,baiocchi2014instrumental,baker2016latent}. Several authors have utilized the IV approach in screening trials for estimating the effect of screening assignment among the compliers, a latent subgroup of participants who adhere to the assigned screening regimen \citep{Baker1998, Roemeling2007}. This is different from the present context where our interest is in the effectiveness of early treatments in the screening detectable subgroup; IV estimators are applicable also for this purpose, and can estimate the mortality risk reduction in the screening detectable subgroup. For this purpose, \citet{mcintosh1999} formulated instrumental variable type estimators in a cohort diagnosed with cancer, assembled after sufficiently long follow-up period to have similar numbers of cancer cases in both screening and control arms to avoid overdiagnosis bias. \citet{saha2018} defined the effect measures and estimators as functions of time, and formulated the estimators in the entire trial cohort, which avoids any selection bias due to conditioning on intermediate variables in instrumental variable analysis \citep{swanson2015}. 

If the screening is discontinued sufficiently early in the trial and the subsequent follow-up is sufficiently long, the estimators proposed by \citet{saha2018} also estimate the case-fatality reduction due to the early treatments, though they are valid for subgroup mortality risk reduction even without this interpretation, which is important if the screening regimen continues for the duration of the trial. The effect measures can be connected to a hypothetical intervention trial proposed by \citet{Miettinen2014, Miettinen2015}, where screening-detected individuals would be randomized into early versus delayed treatments. Although such a trial could not be implemented in practice, it is helpful in formulating meaningful causal estimands that can then be estimated from conventional screening trials. In a different context, treatment effect in a latent subgroup has been studied by \citet{Altstein2011,Altstein2013}, in a diagnostic setting where immediate lymph node removal surgery is received by patients with a positive biopsy test while those with a negative biopsy test receive a control regimen of observation, with the causal effect of the immediate surgery among the positive biopsy subgroup as the quantity of interest. The screening trial context differs from this due to the repeated rounds of screening, making the latent subgroup membership time-dependent.

Recent methodological literature has seen several proposals for instrumental variable estimation with time-to-event and competing risks outcomes. The latter is relevant for the screening context as the interest is in cancer-specific mortality. Instrumental variable estimators generalize to time-to-event outcomes most straightforwardly by considering time-specific risk or survival probability as the outcome and using plug-in estimators that can accommodate censoring such as the Kaplan-Meier estimator \citep{nie2011inference}. This approach generalizes to competing risks outcomes by using non-parametric cumulative incidence estimators \citep{richardson2017nonparametric}. Instrumental variable approaches to estimate causal effect parameters in structural proportional hazards \citep{martinussen2019instrumental} and additive hazards \citep{zheng2017instrumental,martinussen2020instrumental} models have also been proposed. The current setting differs from these because the latent subgroup is time-dependent, accumulated through repeated screening examinations. \citet{saha2018} demonstrated that instrumental variable estimation is still applicable in this setting, formulating the estimators under competing risks, and allowing for inclusion of covariates to account for covariate dependent censoring. 

While \citet{saha2018} estimated absolute and proportional subgroup mortality risk reductions, here we are interested in measuring the early treatment effect in terms of a mortality hazard ratio. For estimation, we broadly follow the approach of \citet{loeys2003}, \citet{loeys2005}, and \citet{martinussen2019instrumental} for instrumental variable estimation for structural Cox proportional hazards models. These authors proposed two-step estimation procedures where estimators for unknown survival probabilities are substituted into an equation derived under the instrumental variable assumptions, to obtain an estimating equation that can be solved with respect to the effect measure of interest. We extend this approach for time-dependent exposure/latent subgroup and competing risks in a multi-state modeling framework. The methods also have similarities to g-estimation in the presence of time-dependent compliance \citep{mark1993method,mark1993estimating,robins1994correcting}. The challenges in formulating causal estimands with time-dependent exposure in a multi-state model setting have been discussed by \citet{von2019causal}. We make sure the quantities are well-defined by linking them with the hypothetical intervention trial.

Because of our focus on the subgroup treatment effect, we aim to characterize randomized screening trials with a simplified multi-state model that avoids modeling the properties of the screening technology (sensitivity, specificity), sojourn time (the period where asymptomatic cancer is screening-detectable), lead time (the period between screening detection and clinical detection in the absence of screening) and stage transition of the cancer. To test the methods, we also use the model as a basis of a data generating mechanism in simulation studies. In contrast, the well-known Hu-Zelen model and its extensions which have been used for planning screening trials \citep{zelen1993,Hu1997,lee2006}, parameter estimation \citep{shen1999parametric,sung2019estimating} and could also serve as a simulation model. These models are based on modeling the natural history of the disease and the properties of the screening test. We aim to avoid specifying these parts of the model by basing our model on the hypothetical intervention trial, where the causal effects of early treatments following an early detection are quantified through a hazard ratio. We will apply the simulation mechanism to compare the performance of the different measures suggested for subgroup mortality reduction, and further use simulation to answer the question of whether conventional screening trials are powered to detect such subgroup effects.

The remainder of the paper is structured as follows: in Section 2 we introduce the necessary notation, define the causal model and the identifying assumptions for estimation under this model and describe the algorithm to simulate data mimicking screening trials. In Section 3, we introduce new estimators to quantify the effect of early treatments in the screening-detectable subgroup as a hazard ratio. We present simulation studies to compare all the estimators in Section 4. In Section 5, we illustrate the use of the new estimator using data from the US National Lung Screening Trial (NLST). Finally, a brief discussion and future directions are presented in Section 6.

\section{A causal model for randomized screening trials}

\subsection{Conventional screening trial}\label{section:conventional}

We introduce the necessary notation in the context of the US National Lung Screening Trial, where $53,452$ heavy smokers, currently smoking or having quit within the last $15$ years, with a smoking history of $30+$ pack years, aged between $55-74$, were assigned at random to undergo three annual screening examinations or control. The participants were asymptomatic of lung cancer at the time of randomization. Participants in the screening arm received three annual low-dose helical CT scans, whereas participants in the control arm received a standard chest X-ray, and both arms were followed up for $7$ years for the main analysis \citep{Team2011}. There were $649$ early diagnoses in the screening arm and $469$ and $552$ cancer deaths in the screening arm and control arm, respectively. In the following, we assume the X-ray screening to have negligible mortality benefits, as has been demonstrated by \citet{Oken2011}, and equate it to a placebo. We characterize the causal effect of screening-induced early treatments through a simplified structural multi-state model where the mortality benefits manifest after an early detection through a CT scan. From the healthy state (numbered as state 1), participants in the screening arm have three possible transitions: early diagnosed through CT screening (state 2), cancer-specific death (state 3) and other-cause death (state 4). From the early diagnosed state, the participants can move to cancer-specific death or other-cause death states, with potentially different transition rates.  We assume that the control arm does not have early diagnosis through CT screening, and thus only direct transitions from state 1 to states 3 and 4 are possible. For concise notation, we use counting processes. We note that since each particular type of transition in a multi-state model can be characterized through a competing risks model, the transition intensities are analogous to the counterfactual cause-specific hazards, as outlined by \citet{young2020causal}, but conditional on the event histories taken place so far.

Let $N^z_{kl}(t) \in \{0,1\}$ be the potential (underlying, in the absence of censoring) counting process that counts the transitions of a given individual from state $k$ to $l$ by time $t$ in arm $z$, where $k \ne l;\; k,l \in \{1,2,3,4\}$ and $z=1$, or $0$ indicate the screening or control arms, respectively. In the current context each transition can take place at most once, and thus each counting process only counts to one, with states 3 and 4 being absorbing states. The corresponding observed process is formulated as $N_{kl}(t)= Z N_{kl}^{1}(t) + (1-Z) N_{kl}^{0}(t)$, where $Z$ is the arm that the patient actually was assigned to in the randomized screening trial (counterfactual consistency). The counting processes are characterized by the transition intensities defined as
\begin{align*}
	\lambda_{kl}^z(t)=\lim_{\Delta t\to 0}\frac{P(N_{kl}^z(t+\Delta t) - N_{kl}^z(t^-)=1 \mid H^z(t^-))}{\Delta t}
\end{align*}

where $H^z(t^-)$ indicates the history of states until time $t^-$. We note that the model could be alternatively specified in discrete time through the conditional probabilities $P(N_{kl}^z(t+\Delta t) - N_{kl}^z(t^-)=1 \mid H^z(t^-)) \approx \lambda_{kl}^z(t)\Delta t$, but we will use continuous time notation for simplicity. The corresponding cumulative transition intensity function is defined as $\Lambda_{kl}^z(t)=\int_{s}^{t} \lambda_{kl}^z(t) \,\textrm dt$. The transition intensities can be collected into a matrix
\[\lambda^z(t) = 
\begin{blockarray}{c@{\hspace{1pt}}rrrr@{\hspace{4pt}}}
	& 1   & 2   & 3 & 4 \\
	\begin{block}{r@{\hspace{1pt}}|@{\hspace{1pt}}
			|@{\hspace{1pt}}rrrr@{\hspace{1pt}}|@{\hspace{1pt}}|}
		1 &  . & \lambda_{12}^z(t) & \lambda_{13}^z(t) & \lambda_{14}^z(t)\\
		2 & 0 &  . & \lambda_{23}^z(t)   & \lambda_{24}^z(t) \\
		3 & 0 & 0 &  . & 0 \\
		4 & 0 &  0 & 0  &  . \\
	\end{block}
\end{blockarray}\]


In the conventional screening trials, the ITS estimand would be defined by comparing the outcomes between the two arms at time $t$ after sufficiently long follow-up, in terms of the cumulative incidences for cancer-specific mortality. For example, the absolute reduction in cancer-specific mortality risk would be measured by  
\begin{equation}\label{equation:itt}
	E[N_{13}^{0}(t)] - E[N_{13}^{1}(t)+N_{23}^{1}(t)]
\end{equation}
and the proportional reduction by
\begin{equation}\label{equation:ittp}
	1 - \frac{E[N_{13}^{1}(t)+N_{23}^{1}(t)]}{E[N_{13}^{0}(t)]}
\end{equation}
The estimation of these quantities does not require a multi-state model, as the components could be estimated through empirical cumulative incidences in the two arms. However, the multi-state model is needed to specify and estimate the causal effect of the early treatment among the early detectable subgroup. This subgroup is latent at baseline and accumulates in the screening arm during the follow-up of the trial through the repeated screening examinations. Since screening itself is not an intervention, the mortality benefits can manifest only after an early detection. To specify the corresponding causal effect, we need to introduce a well-defined intervention.

\subsection{Hypothetical intervention trial}

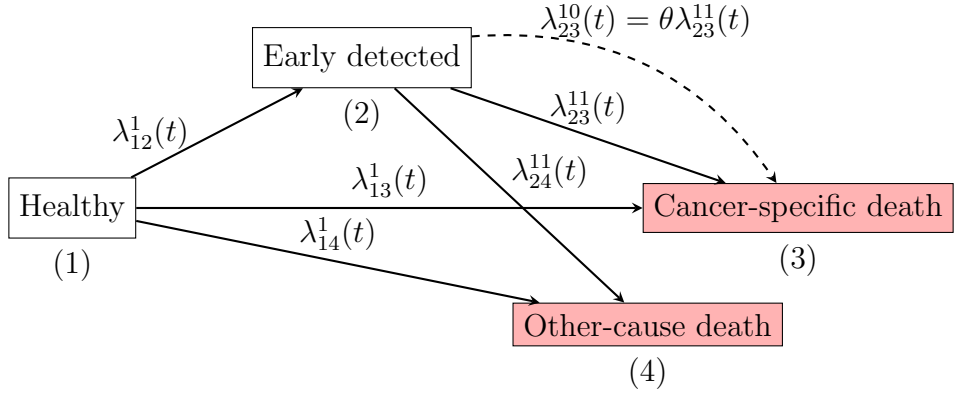
\begin{figure}[!ht]
	\begin{center}
		\begin{tikzpicture} [node distance=1.15cm]
			\node(scr) [inter,label=below:(1)] {Healthy};
			\node(Cancer) [inter,right of=scr,xshift=2.7cm,yshift=2cm,label=below:(2)] {Early detected};
			\node(CD) [startstop,right of=scr,xshift=8.5cm,label=below:(3)] {Cancer-specific death};
			\node(OD) [startstop, right of=scr, xshift=6.5cm,yshift=-1.55cm,label=below:(4)] {Other-cause death};
			\draw [arrow] (scr) -- node [label={[label distance=0.07cm](180):\textcolor{black}{$\lambda_{12}^1(t)$}}]{}  (Cancer);
			\draw [arrow] (scr) -- node [anchor=south] {$\lambda_{13}^1(t)$} (CD);
			\draw [arrow] (scr) -- node [anchor=south] {$\lambda_{14}^1(t)$} (OD);
			\draw [arrow] (Cancer) -- node [anchor=south] {$\lambda_{23}^{11}(t)$} (CD);
			\draw [arrow] (Cancer) -- node [label={[label distance=-0.3cm](20):\textcolor{black}{$\lambda_{24}^{11}(t)$}}]{} (OD);
			\path [arrow] (Cancer) edge [bend left=30,dashed] node [label={[label distance=0.07cm](90):\textcolor{black}{$\lambda_{23}^{10}(t) = \theta \lambda_{23}^{11}(t)$}}] {} (CD);
		\end{tikzpicture}
		\caption{A schematic illustrating the transition intensities in the hypothetical intervention trial. Here $\lambda_{kl}(t)$ corresponds to the transition intensity from state $k$ to state $l$ at time $t$. The dashed arrow represents cancer mortality under delayed treatments, contrasted to early treatments, with the intensity ratio $\theta$ quantifying the causal effect of interest.}
		\label{fig:multi_state}
	\end{center}
\end{figure}

Following and extending \citet{Miettinen2015}, we can envision a trial that has similar screening regimen as the screening arm of the conventional trial, but instead of randomizing into screening and non-screening, screens everyone, and at the time of an early detection, randomizes into referral to early treatment vs delayed treatment through withholding the screening result. Delayed treatment refers treatments following a later symptomatic diagnosis. Since the early detection can take place only once for a given individual, the randomization for the assignment in the event of early detection can already take place at study baseline, which is relevant for the indexing of the corresponding potential outcomes. A schematic illustration of the hypothetical trial is presented in Figure \ref{fig:designs}, and the corresponding multi-state model in Figure \ref{fig:multi_state}. While this kind of trial could not be implemented in practice, it is important as a thought experiment to define the causal quantities that we are interested in. 

Among $N_{12}^{1}(t)=1$ we define $N^{1r}_{2l}(t) \in \{0,1\}$, $l \in \{3,4\}$ to be potential (underlying) counting processes for subsequent death corresponding to early ($r=1$) versus delayed ($r=0$) treatment, characterized by transition intensities 
\begin{align*}
	\lambda_{2l}^{1r}(t)=\lim_{\Delta t\to 0}\frac{P(N_{2l}^{1r}(t+\Delta t) - N_{2l}^{1r}(t^-)=1 \mid H^1(t^-))}{\Delta t}.
\end{align*} 
The modified transition matrix for screening detectable subgroup is now
\[\lambda^{1r} (t) = 
\begin{blockarray}{c@{\hspace{1pt}}rrrr@{\hspace{4pt}}}
	& 1   & 2   & 3 & 4 \\
	\begin{block}{r@{\hspace{1pt}}|@{\hspace{1pt}}
			|@{\hspace{1pt}}rrrr@{\hspace{1pt}}|@{\hspace{1pt}}|}
		1 &  . & \lambda_{12}^{1}(t) & \lambda_{13}^{1}(t) & \lambda_{14}^{1}(t)\\
		2 & 0 &  . & \lambda_{23}^{1r}(t)   & \lambda_{24}^{1r}(t) \\
		3 & 0 & 0 &  . & 0 \\
		4 & 0 &  0 & 0  &  . \\
	\end{block}
\end{blockarray}\]

Under this model, the absolute cancer mortality risk reduction in the subgroup could be measured by
\begin{equation}\label{equation:acfr}
	E[N_{23}^{10}(t) - N_{23}^{11}(t)  \mid N_{12}^{1}(t)=1]
\end{equation}
and the proportional reduction by
\begin{equation}\label{equation:pcfr}
	1-\frac{E[N_{23}^{11}(t)=1 \mid N_{12}^{1}(t)=1]}{E[N_{23}^{10}(t)=1 \mid N_{12}^{1}(t)=1]}.
\end{equation}
We note that these quantities are distinct from \eqref{equation:itt} and \eqref{equation:ittp} as they are defined in the early diagnosed subpopulation. Conditioning on this subgroup membership also conditions on survival until the early diagnosis, so \eqref{equation:acfr} and \eqref{equation:pcfr} measure mortality reductions after the early diagnosis. In a trial where the screening is discontinued at some point (as it was in NLST after three annual rounds), and the follow-up is long enough so that all the mortality benefits have been realized, these quantities approximate the absolute and proportional case-fatality reduction, contrasting the screening-induced early versus delayed treatment among the screening detectable cases. The estimation of these quantities was discussed by \citet{saha2018}. 

Herein we aim to measure the effect by contrasting the transition intensities $\lambda_{23}^{11}(t)$ and $\lambda_{23}^{10}(t)$. We note that the formulation in terms of transition intensities does not yet assume anything about the early treatment effect, for example generally this effect could be a function of time and/or function of the time of the early diagnosis. However, we can simplify the model by characterizing the effect as a constant ratio $\lambda_{23}^{10}(t) = \theta \lambda_{23}^{11}(t)$, characterizing the early treatment effect on cancer-specific mortality through a single parameter, corresponding to the proportionality of the two cause-specific mortality hazards. In addition, although this is not generally required, we take $\lambda_{24}^{10}(t) = \lambda_{24}^{11}(t)$, meaning that the early treatments do not have effect on other-cause mortality. In what follows, we focus on the estimation of the quantity $\theta$ as our effect measure of interest. We note that we parametrized this as the ratio of the mortality under delayed vs. early treatment to be consistent with the earlier defined risk reduction measures. Alternatively, the inverse of this can be reported. Herein we take the time scale for all of the transition intensities to be the time since the baseline/study entry. However, it would also be equally possible to use the time since early detection as the time scale for the transitions after the early detection, and to define proportionality of the effect on this time scale.

Since the transition intensities $\lambda_{2l}^{10}(t)$ are not directly estimable from data observed under a conventional randomized screening trial, the estimation requires instrumental variable type methods, using the randomized assignment in the conventional trial as an instrument. We will propose estimation methods in Section \ref{section:methods}, but will first discuss the use of the model for simulation, and outline the assumptions required for the instrumental variable estimation.

\begin{figure}[h!]
	\centering
	\resizebox{1\linewidth}{!}{
		\begin{tikzpicture}[node distance=3.03cm, text width=2cm]
			\footnotesize
			\node (pop) [startstop2] {Eligible for screening and randomized};
			
			\node (z1) [startstop2,below of=pop,xshift=-1.9cm] {$Z=1$\\(screening arm)};
			
			\node (z0) [startstop2, below of=pop, xshift=1.9cm] {$Z=0$\\(control arm)};
			
			\node (y1) [startstop2, below of=z1] {$N_{13}^{1}(t)+N_{23}^{1}(t)$};
			\node (y0) [startstop2, below of=z0] {$N_{13}^{0}(t)$};
			\node at (0,-7.5,1) [text width=6.5cm, align=left] {(a) Conventional screening trial\\$\quad\;\;$Contrast: $1 - \frac{E[N_{13}^{1}(t)+N_{23}^{1}(t)]}{E[N_{13}^{0}(t)]}$ \\$\quad\;\;$or $E[N_{13}^{0}(t)] - E[N_{13}^{1}(t)+N_{23}^{1}(t)]$};
			\draw [arrow] (pop) -- (z1);
			\draw [arrow] (pop) -- (z0);
			\draw [arrow] (z1) -- (y1);
			\draw [arrow] (z0) -- (y0);
			\node (pop2) [startstop2,right of=pop,xshift=4.1cm] {Eligible for screening};
			\node (z1) [startstop2,below of=pop2,yshift=1.55cm] {$Z=1$\\(everyone screened)};
			\node (d1) [startstop2,below of=z1,yshift=1.35cm] {$N_{12}^{1}(t)=1$\\(screening-detected and randomized)};
			\node (r1) [startstop2,below of=d1,xshift=-1.9cm,yshift=1.5cm] {$R=1$ (referral to early treatment)};
			\node (r0) [startstop2,below of=d1,xshift=1.9cm,yshift=1.5cm] {$R=0$ (no referral to early treatment)};
			\node (y11) [startstop2,below of=r1,yshift=1.65cm] {$N_{23}^{11}(t)$};	
			\node (y10) [startstop2,below of=r0,yshift=1.65cm] {$N_{23}^{10}(t)$};
			\draw [arrow] (pop2) -- (z1);
			\draw [arrow] (z1) -- (d1);
			\draw [arrow] (d1) -- (r1);
			\draw [arrow] (d1) -- (r0);
			\draw [arrow] (r1) -- (y11);
			\draw [arrow] (r0) -- (y10);
			\node at (8,-7.75,1) [text width=6.5cm] {(b) Hypothetical intervention trial\\
				$\quad\;\;$Contrast: $1- \frac{E[N_{23}^{11}(t) \mid N_{12}^{1}(t)=1]}{E[N_{23}^{10}(t) \mid N_{12}^{1}(t)=1]}$ \\$\quad\;\;$or $E[N_{23}^{10}(t) - N_{23}^{11}(t) \mid N_{12}^{1}(t)=1]$};
	\end{tikzpicture}}
	\caption{Illustration of a conventional randomized screening trial and a hypothetical intervention trial sharing the same screening regimen, and the corresponding causal contrasts.}
	\label{fig:designs}
\end{figure}

\subsection{Identifying assumptions}\label{section:assumptions}

In the methods proposed in Section \ref{section:methods}, the random assignment in the conventional screening trial is used as an instrumental variable. We outline the identifying assumptions here as they are also used to generate simulated data from the model. In particular, we assume that (i) $N_{12}^1(t) = 1 \Rightarrow N_{1l}^0(t)= N_{2l}^{10}(t)$, $l \in \{3,4\}$, meaning that for someone who would get early detected by time $t$, but not receive early treatment, the potential mortality outcome would be the same as in the control arm of the conventional trial. Further, we assume that (ii) $N_{12}^1(t) = 0 \Rightarrow N_{1l}^0(t) = N_{1l}^1(t)$, $l \in \{3,4\}$, meaning that for someone who would not get early detected, the potential mortality outcome would be the same between the two arms of the conventional trial. Essentially, assumptions (i) and (ii) together state that screening examinations themselves do not have impact on mortality in the absence of the intervention. They broadly correspond to the exclusion restriction assumption of conventional instrumental variable analysis \citep{angrist1996}, stating that the instrumental variable does not have direct causal effect on the outcome, outside the effect mediated by the intervention. In the present context this assumption could be violated for example if the screening examinations cause the participants to modify their health behaviour in a way that affects their cancer risk. In the NLST also the control arm received screening, so any such effects should be more similar between the two arms, making (i) and (ii) more plausible. 

In addition, we assume that (iii) $N_{12}^{1}(t)=1 \Rightarrow N_{2l}^{1}(t) = N_{2l}^{11}(t)$, $l \in \{3,4\}$, meaning that in practice in the screening arm early treatment follows early diagnosis, which is reasonable at least in the lung cancer context. From this it follows that $\lambda_{2l}^{1}(t) = \lambda_{2l}^{11}(t)$, meaning that the mortality rates under early treatment following early diagnosis can be estimated from the screening arm of the conventional trial. For the purpose of identification based on data from the two arms of the conventional trial, we also need counterfactual consistency (iv), formulated in Section \ref{section:conventional}. Finally, we take that $\lambda_{12}^0(t)=\lambda_{23}^0(t)= \lambda_{24}^0(t)=0$, meaning that early detection through the screening technology offered in the screening arm is not available in the control arm.

\subsection{Description of the simulation algorithm}\label{section:algorithm}

To simulate data from the proposed multi-state model, one needs to fix the transition intensities in the matrix $\lambda^{1r}(t)$, to simulate event histories in the two arms in the hypothetical intervention trial, corresponding to $r=1$ and $r=0$. These event histories are also used to represent event histories in the conventional screening trial. In particular, under condition (iii) the intervention arm ($r=1$) event histories can be directly taken to be event histories in the screening arm of the conventional trial. Also, the control arm $r=0$ event histories can be directly taken to be event histories in the control arm by taking $N_{12}^{0}(t) = 0$ and $N_{1l}^{0}(t) = N_{1l}^{1}(t) + N_{2l}^{10}(t)$, $l \in \{3,4\}$, which follows from (i) and (ii). If the outcomes are simulated without covariates, the data generating mechanism and the structural model are the same. However, for the data generation, all the intensities can be made conditional on observed and/or unobserved simulated covariates, in which case the structural model will parametrize the marginal effect of interest. In this case, the marginal effect is not directly specified by the data generating mechanism, but can be approximated by fitting a marginal proportional hazards model to the two arms of the hypothetical trial.

For a simulation study, for example exponential or Weibull functional forms can be assumed for the intensities, or the observable ones ($\lambda_{1l}^{1}(t)$, $l \in \{2,3,4\}$ and $\lambda_{2l}^{11}(t)$, $l \in \{3,4\}$) can be estimated from the screening arm of an existing trial such as the NLST, with the control arm ($r = 0$) counterparts obtained by fixing the effect parameter $\theta$ to a value. With the transition intensity matrix $\lambda^{1r}$, simulation can then proceed through usual algorithms for simulating event histories from a multi-state model as described for example in \citet[p.~ 45]{beyersmann2011}. Briefly, this proceeds by simulating the time and type of the next event from a competing risks model, moving to a new state, and continuing the same. Starting from the healthy state, this would mean simulating the time of the first event, $T$, from the total hazard
\begin{align*}
	\Lambda^1(t)= \Lambda_{12}^1(t)+\Lambda_{13}^1(t)+\Lambda_{14}^1(t),
\end{align*}
which specifies the event time distribution $F(t)=P(T\le t)=1-\exp(-\Lambda^1(t))$. Using the inverse cumulative distribution function method, we can take $u \sim U[0,1]$ and solve $\Lambda^1(t)+\log(1-u)=0$ with respect to $t$ to get the time. At time $T=t$ the type of the event is then drawn with multinomial probabilities
\begin{align*}
	\frac{\lambda_{1l}^1(t)}{\lambda_{12}^1(t)+\lambda_{13}^1(t)+\lambda_{14}^1(t)}
\end{align*}
for $l \in \{2,3,4\}$. If the new state is terminal ($l \in \{3,4\}$), the algorithm stops, otherwise it continues similarly from state 2 after possible modification of the subsequent transition intensities based on the timing of the first event, where the new total hazard is given by $\Lambda^{1r}(t)= \Lambda_{23}^{1r}(t)+\Lambda_{24}^{1r}(t)$, and the event type probabilities by
\begin{align*}
	\frac{\lambda_{2l}^{1r}(t)}{\lambda_{23}^{1r}(t)+\lambda_{24}^{1r}(t)}
\end{align*}
for $l \in \{3,4\}$.
We used the NLST as the basis of the simulation study presented in Section \ref{section:simulation} by estimating the quantities $\lambda^{11} (t)$ from the screening arm of the trial. For the observable transition intensities, we calculated smooth hazard estimates using the \texttt{muhaz} \citep{muhaz} package in R and integrated them numerically to get the corresponding cumulative intensities.

\section{Hazard ratio estimation}\label{section:methods}

\subsection{Estimating equation}\label{section:equation}

To derive a connection between the observable and unobservable quantities under the identifying assumptions listed in Section \ref{section:assumptions}, we can re-express the cumulative incidence of cancer-specific mortality in the control arm of a conventional trial $E[N_{13}^0(t)]$ as the sum of two different alternative event histories in terms of the latent subgroup membership of the individual, that is, the event history under early detection and no early treatment and the event history under no early detection by time $t$. This gives
\begin{align*}
	E[N_{13}^0(t)] &=  \int_{0}^{t} \exp \left\{-(\Lambda_{13}^0(u)+\Lambda_{14}^0(u))\right\} \lambda_{13}^0(u) \,\textrm du \nonumber \\
	&= E[N_{13}^{0}(t) \mid N_{12}^1(t)=1] P(N_{12}^1(t)=1) \nonumber \\
	&\quad+ E[N_{13}^0(t) \mid N_{12}^1(t)=0]P(N_{12}^1(t)=0) \nonumber \\
	&\stackrel{\textrm{(i),(ii)}}{=} E[N_{23}^{10}(t) \mid N_{12}^1(t)=1] P(N_{12}^1(t)=1) \nonumber \\
	&\quad+  E[N_{13}^1(t) \mid N_{12}^1(t)=0] P(N_{12}^1(t)=0) \nonumber \\
	&= \int_{0}^{t} \exp \left\{-(\Lambda_{12}^{1}(u)+\Lambda_{13}^{1}(u)+\Lambda_{14}^1(u)) \right\} \lambda_{12}^{1}(u) \nonumber\\
	&\quad\times \int_{u}^{t} \exp \left\{- \left(\int_{u}^{s} [\lambda_{23}^{10}(s) + \lambda_{24}^{10}(s)] \,\textrm ds\right) \right\} \lambda_{23}^{10}(v) \,\textrm dv\,\textrm du \nonumber\\
	&\quad+ \int_{0}^{t} \exp \left\{-(\Lambda_{12}^1(u)+\Lambda_{13}^1(u)+\Lambda_{14}^1(u))\right\} \lambda_{13}^1(u) \,\textrm du
\end{align*}
Here the second equality used assumptions (i) and (ii). Finally, applying assumption (iii), as well as the modeling assumption of proportional hazards, we get
\begin{align}\label{equation:cumulative_control_cancer}
	E[N_{13}^0(t)]
	&\stackrel{\textrm{(iii)}}{=} \int_{0}^{t} \exp \left\{-(\Lambda_{12}^{1}(u)+\Lambda_{13}^{1}(u)+\Lambda_{14}^1(u)) \right\} \lambda_{12}^{1}(u) \nonumber\\
	&\quad\times \int_{u}^{t} \exp \left\{- \left(\int_{u}^{v} [\theta\lambda_{23}^{1}(s) + \lambda_{24}^{1}(s)] \,\textrm ds\right) \right\} \theta \lambda_{23}^{1}(v) \,\textrm dv\,\textrm du \nonumber\\
	&\quad+ \int_{0}^{t} \exp \left\{-(\Lambda_{12}^1(u)+\Lambda_{13}^1(u)+\Lambda_{14}^1(u))\right\} \lambda_{13}^1(u) \,\textrm du
\end{align}
A corresponding equation using time since the early diagnosis as the time scale after transition to state 2 would be obtained by replacing $\lambda_{2l}^{1}(v)$ with $\lambda_{2l}^{1}(v-u)$, $l \in \{3,4\}$. A related equation was considered by \citet{mcintosh1999}, but without considering the hazard ratio as the parameter of interest. We note that in the case of only single baseline screening examination at time $t=0$ and the absence of competing risks, \eqref{equation:cumulative_control_cancer} would simplify to
\begin{align*}
	E[N_{13}^0(t)] &= E[N_{23}^{10}(t) \mid N_{12}^1(0)=1] P(N_{12}^1(0)=1) \\
	&\quad+ E[N_{13}^1(t) \mid N_{12}^1(0)=0]P(N_{12}^1(0)=0) \\
	&=P(N_{12}^1(0)=1) \int_{0}^{t} \exp \left\{-\theta \Lambda_{23}^{1}(u) \right\}\theta \lambda_{23}^{1}(u) \, \textrm du \\
	&\quad+P(N_{12}^1(0)=0) \int_{0}^{t} \exp \left\{-\Lambda_{13}^{1}(u) \right\} \lambda_{13}^{1}(u) \, \textrm du,
\end{align*} 
which corresponds to the equality given by \citet[][p. 69]{martinussen2019instrumental}. 

In addition to the parameter of interest $\theta$, \eqref{equation:cumulative_control_cancer} involves other unknown quantities, needing to be estimated. We note that under counterfactual consistency (iv) all the quantities (except $\theta$) on the right hand side of \eqref{equation:cumulative_control_cancer} can be estimated from the screening arm of the conventional trial, while the left hand size can be estimated from the control arm of the conventional trial. In principle, any associational model can be used to obtain estimators to substitute in for the quantities on the right hand side. For example, proportional cause-specific hazard models where the mortality after the early diagnosis is allowed to depend on time $u$ of the early diagnosis can be fitted for $\lambda_{2l}^{1}(v)$, $l \in \{3,4\}$. However, here we focus on the special case of Markov multi-state model, where $\lambda_{2l}^{1}(v)$ does not depend on $u$, in which case we can use non-parametric estimators to substitute in for the unknown quantities.

State occupation probabilities in a Markov multi-state model can be generally estimated using the non-parametric Aalen-Johansen approach \citep{borgan2014alen}, which in the case of a competing risks setting reduces to the non-parametric cumulative incidence estimator for the left hand side of the equation. The required inputs for the right hand side are Nelson-Aalen/Breslow estimates for the cumulative hazards/hazard increments; using these as inputs (with one of them modified by the multiplicative factor $\theta$), the Aalen-Johansen state occupation probabilities can be calculated for example using the \texttt{msfit} and \texttt{probtrans} functions of the \texttt{mstate} package \citep{putter:2007,mstate}. Equation \eqref{equation:cumulative_control_cancer} can then be solved numerically with respect to $\log\theta$ to estimate the early treatment effect. 

Equation \eqref{equation:cumulative_control_cancer} was written for the underlying counting processes in the absence of censoring. However, by substituting in estimators that can accommodate independent censoring (unconditionally on covariates), the proposed approach can accommodate independent censoring. We discuss in Section \ref{section:discussion} how covariate-dependent censoring can be accommodated. The fixed time $t$ at which the estimating equation is evaluated can be chosen as the maximum follow-up length, or, as we demonstrate in Section \ref{section:illustration}, the effect can be estimated as a function of $t$ to see if the estimates stabilize over the follow-up period. Alternatively, if reporting a single estimate is desirable, the timepoint for the estimation can be chosen to minimize the empirical variance as $\hat\theta = \arg\min_t \hat V(\hat\theta_t)$ as suggested by \citet{martinussen2019instrumental}, or as the inverse variance weighted average $\hat\theta = \left(\sum_t \hat\theta_t/\hat V(\hat\theta_t)\right)/\left(\sum_t 1/\hat V(\hat\theta_t)\right)$ over an equally spaced grid of times $t$. The variance estimates can be obtained through the bootstrap.

We note that we can obtain an analogous equation 
\begin{align}\label{equation:cumulative_control_other}
	E[N_{14}^0(t)] &= \int_{0}^{t} \exp \left\{-(\Lambda_{12}^{1}(u)+\Lambda_{13}^{1}(u)+\Lambda_{14}^1(u)) \right\} \lambda_{12}^{1}(u) \nonumber\\
	&\quad\times \int_{u}^{t} \exp \left\{- \left(\int_{u}^{v} [\theta \lambda_{23}^{1}(s) + \lambda_{24}^{1}(s)] \,\textrm ds\right) \right\} \lambda_{24}^{1}(v) \,\textrm dv\,\textrm du \nonumber\\
	&\quad+ \int_{0}^{t} \exp \left\{-(\Lambda_{12}^1(u)+\Lambda_{13}^1(u)+\Lambda_{14}^1(u))\right\} \lambda_{14}^1(u) \,\textrm du.
\end{align}
for other-cause mortality. However, this is not needed for estimation as we have only one unknown quantity to solve under the assumption that early treatments do not have effect on other-cause mortality. We note that the proposed methodology can be extended to relax this assumption, that is $\lambda^{10}_{24}(t) = \lambda^{11}_{24}(t)$. In that case, we can solve the two equations for two unknowns, characterizing the effect on cancer-specific and other cause mortality, respectively.

\subsection{Maximum likelihood estimation}\label{section:likelihood}

Alternatively to the estimating equation approach, we can consider a multinomial likelihood expression for the parameter $\theta$ using data from the mortality outcomes in the conventional trial. This can be written as 
\begin{align}\label{equation:lik}
	L(\theta) &= \prod_{\{i:Z_i = 1\}} \Big\{ {E[N_{13}^1(t) + N_{23}^1(t)]}^{N_{13}^1(t)+ N_{23}^1(t)}  {E[N_{14}^1(t) + N_{24}^1(t)]}^{N_{14}^1(t)+N_{24}^1(t)} \nonumber \\
	&\qquad\times \left(1-E[N_{13}^1(t) + N_{23}^1(t)]-E[N_{14}^1(t) + N_{24}^1(t)]\right)^{1-N_{13}^1(t)-N_{23}^1(t)-N_{14}^1(t) - N_{24}^1(t)} \Big\} \nonumber \\  
	&\quad\times \prod_{\{i:Z_i = 0\}} E[N_{13}^0(t)]^{N_{13}^{0}(t)}  E[N_{14}^0(t)]^{N_{14}^{0}(t)} \nonumber \\
	&\qquad\times \left(1-E[N_{13}^0(t)]-E[N_{14}^0(t)]\right)^{(1-N_{13}^0(t)-N_{14}^0(t))} \nonumber \\
	&\stackrel{\theta}{\propto} \prod_{\{i:Z_i = 0\}} E[N_{13}^0(t)]^{N_{13}^{0}(t)}  E[N_{14}^0(t)]^{N_{14}^{0}(t)} \nonumber \\
	&\qquad\times \left(1-E[N_{13}^0(t)]-E[N_{14}^0(t)]\right)^{1-N_{13}^0(t)-N_{14}^0(t)},
\end{align}
where $E[N_{13}^0(t)]$ and $E[N_{14}^0(t)]$ are as in \eqref{equation:cumulative_control_cancer} and \eqref{equation:cumulative_control_other}. The proportionality followed because, considering the transition intensities in the screening arm as fixed quantities, the likelihood contributions of the individuals in the screening arm do not depend on $\theta$. Similarly to the estimating equation in Section \ref{section:equation}, \eqref{equation:lik} involves unknown quantities that under the stated assumptions can be replaced with estimators from the screening arm of the trial, and the expression $L(\theta)$ then maximized numerically with respect to $\log \theta$. The limitation of the multinomial likelihood expression \eqref{equation:lik} is that it can only accommodate type I censoring. However, instead of evaluating the likelihood expression at a fixed time $t$, independent censoring can be accommodated by evaluating the likelihood contributions at the minimum of the time of death or censoring time. Such a likelihood expression could also be informative of time-dependent treatment effects. We will briefly outline this in Section \ref{section:discussion}.

\section{Simulation studies}

\subsection{Using NLST as basis}\label{section:simulation}

To compare the estimators proposed in Section \ref{section:methods} in a setting resembling a real randomized screening trial, we generated data using the algorithm described in Section \ref{section:algorithm}, using the NLST as the basis of the simulation study. In particular, because the NLST was powered for the ITS effect, we were interested in determining if such a study is also powered for detecting the subgroup effect using tests based on the proposed estimators. For the purpose of comparison, we also included tests based on these to the two estimators proposed earlier for absolute and proportional subgroup mortality risk reduction \citep{saha2018}. In the present notational framework, the absolute mortality risk reduction \eqref{equation:acfr} can be expressed as
\begin{align*}
	\MoveEqLeft E[N_{13}^{0}(t)- \{N_{13}^{1}(t)+N_{23}^{1}(t)\}] \\
	&= E[N_{13}^{0}(t)- \{N_{13}^{1}(t)+N_{23}^{1}(t)\} \mid N_{12}^{1}(t) = 1] P(N_{12}^{1}(t) = 1) \\
	&\quad+E[N_{13}^{0}(t)- \{N_{13}^{1}(t)+N_{23}^{1}(t)\}\mid N_{12}^{1}(t) = 0] P(N_{12}^{1}(t) = 0) \\
	&\stackrel{\textrm{(i),(ii),(iii)}}{=} E[ N_{23}^{10}(t)- \{N_{13}^{1}(t)+N_{23}^{11}(t)\} \mid N_{12}^{1}(t) = 1] P(N_{12}^{1}(t) = 1)\\
	&=E[N_{23}^{10}(t)- N_{23}^{11}(t) \mid N_{12}^{1}(t) = 1] E[N_{12}^{1}(t)]
\end{align*}
Thus, under (iv) we get,
\begin{align}\label{equation:ACFR}
	\eqref{equation:acfr} 
	= \frac{E[N_{13}(t) \mid Z = 0]-E[N_{13}(t)+N_{23}(t) \mid Z = 1]}{E[N_{12}(t) \mid Z = 1]},
\end{align}
where the expectations can be estimated non-parametrically as cumulative incidences. Further, we can write
\begin{align*}
	\MoveEqLeft E[N_{13}^0(t)] \\
	&=E[N_{13}^0(t) \mid N_{12}^1(t)=1] P(N_{12}^1(t)=1) \\
	&\quad+E[N_{13}^0(t) \mid N_{12}^1(t)=0] P(N_{12}^1(t)=0) \\
	&\stackrel{\textrm{(i),(ii)}}{=} E[N_{23}^{10}(t) \mid N_{12}^1(t)=1] P(N_{12}^1(t)=1) \\
	&\quad\quad+E[N_{13}^{1}(t) \mid N_{12}^1(t)=0] P(N_{12}^1(t)=0)\\
	&= E[N_{23}^{10}(t) \mid N_{12}^1(t)=1] P(N_{12}^1(t)=1) + P(N_{13}^{1}(t)=1, N_{12}^1(t)=0)\\
	&= E[N_{23}^{10}(t) \mid N_{12}^1(t)=1] P(N_{12}^1(t)=1) + E[N^1_{13}(t)]
\end{align*}
Using the above two results, we can now re-express \eqref{equation:pcfr} as,
\begin{align*}
	\MoveEqLeft 1-\frac{E[N_{23}^{11}(t) \mid N_{12}^{1}(t)=1]}{E[N_{23}^{10}(t) \mid N_{12}^{1}(t)=1]} \\
	&= \frac{(E[N_{23}^{10}(t) \mid N_{12}^{1}(t)=1]-E[N_{23}^{11}(t) \mid N_{12}^{1}(t)=1])P(N_{12}^1(t)=1)}{E[N_{23}^{10}(t)=1 \mid N_{12}^{1}(t)=1]P(N_{12}^1(t)=1)} \\
	&=\frac{E[N^0_{13}(t)]-E[N^1_{13}(t)+N^1_{23}(t)]}{E[N_{13}^0(t)] - E[N^1_{13}(t)]}
\end{align*} 
Finally, under (iv) we can write that
\begin{align}\label{equation:PCFR}
	\eqref{equation:pcfr} 
	=\frac{E[N_{13}(t) \mid Z = 0] - E[N_{13}(t) + N_{23}(t) \mid Z = 1]}{E[N_{13}(t) \mid Z = 0] - E[N_{13}(t) \mid Z = 1]},
\end{align}
where again the expectations can be replaced with appropriate non-parametric cumulative incidence estimators. While \eqref{equation:ACFR} and \eqref{equation:PCFR} are estimating mortality risk reductions rather than the hazard ratio, the four estimators are comparable in terms of the power of a Wald-type or confidence interval based test on them. This will help answer the question of whether conventional screening trial such as NLST is powered to detect the subgroup early treatment effect. For this reason, we also calculated estimates for the ITS mortality reductions as
\begin{equation}\label{equation:ITT}
	\eqref{equation:itt} 
	=E[N_{13}(t) \mid Z=0] - E[N_{13}(t)+N_{23}(t) \mid Z=1]
\end{equation}
and the proportional reduction by
\begin{equation}\label{equation:ITTP}
	\eqref{equation:ittp} 
	=1 - \frac{E[N_{13}(t)+N_{23}(t) \mid Z=1]}{E[N_{13}(t) \mid Z=0]},
\end{equation}
as the trial is powered for these.

To compare the four estimators, we generated $1000$ NLST-like datasets, by fixing the transition intensities to those estimated from the NLST, and fixing the $\log$ hazard ratio to $0.4804$, estimated from the real data using the estimating equation \eqref{equation:cumulative_control_cancer}. While we could have fixed the hazard ratio to any value in the simulation, we used the empirical value to obtain similar death counts to the real data. For each dataset we simulated $n=53,452$ individuals assigned randomly with equal probability to screening or control. Since in the NLST the censoring is mostly administrative at the end of the follow-up at 7 years, we did not simulate random censoring, so the simulated datasets had only type I censoring at 7 years. The true values of the subgroup mortality reductions were evaluated by calculating the quantities \eqref{equation:acfr} and \eqref{equation:pcfr} under the specified transition intensities, that is, evaluating the cumulative incidences through numerical integration using the true transition intensity functions used in the simulation. In the estimators, the cumulative incidences were calculated using the \texttt{cuminc} function of the \texttt{cmprsk} R package \citep{cmprsk}. For more complex state occupation probabilities needed for evaluating the estimating equation and the likelihood expression, we first calculated Nelson-Aalen/Breslow estimators, which were used as inputs to Aalen-Johansen estimators for state occupation probabilities using the functions available in the \texttt{mstate} R package. The estimating equation was solved numerically using the \texttt{uniroot} function and the likelihood maximized numerically using the \texttt{optim} function of R \citep{rmanual}.

The mortality reductions, as well as the estimating equation \eqref{equation:cumulative_control_cancer} and the likelihood expression \eqref{equation:lik} were evaluated at 7 years. To estimate standard errors for each estimator, we bootstrapped each simulated dataset 50 times, and took the standard deviation of the point estimates as the standard error. This was compared to the Monte Carlo standard deviation of the point estimates over the simulation rounds. 95\% confidence intervals were constructed using the normal approximation and compared to the nominal coverage probability. The power of the confidence interval based test was calculated as the proportion of simulation rounds where the interval did not cover the null value. 

The simulation results are presented in Table \ref{table}. These indicate that all the estimators could capture the corresponding true value relatively well, based on low bias and near nominal coverage probability. The power of the subgroup treatment effect estimators was also comparable to the ITS analysis. This is consistent with the similar property of g-estimation methods \citep{white2005uses}. This can be explained by the fact that all the mortality reduction due to screening is realized through the effect of the early treatments, and thus focusing on the subgroup does not lose any of the effect. The test based on the estimator \eqref{equation:PCFR} had the highest power. The two hazard ratio estimators gave very similar results, so we only use the estimating equation for the data analysis in the next section.

\begin{table}[ht!]
	\centering
	\begin{tabular}{lllllllll}
		\hline
		\multirow{2}{*}{Estimator}  & Causal & \multirow{2}{*}{Truth} & Point & \multirow{2}{*}{SE} & \multirow{2}{*}{Power} & \multirow{2}{*}{Coverage} & \multirow{2}{*}{MCsd} &\multirow{2}{*}{MCE} \\
		& contrast & & estimate & & & & & \\
		\hline
		\eqref{equation:cumulative_control_cancer} & $\log \theta$ & 0.4804 & 0.4893  & 0.1628 & 0.85 & 0.942 & 0.1657 & 0.005 \\
		\eqref{equation:lik} & $\log \theta$ & 0.4804 & 0.4892 & 0.1628 & 0.85 & 0.942 & 0.1657 & 0.005\\
		\eqref{equation:ACFR} & \eqref{equation:acfr} & 0.1512 & 0.1507 & 0.0521 & 0.83 & 0.945 & 0.0534 & 0.0017 \\
		\eqref{equation:PCFR} & \eqref{equation:pcfr} & 0.2986 & 0.2921 & 0.0804 & 0.90 & 0.944 & 0.0803 & 0.0025\\
		\eqref{equation:ITT} & \eqref{equation:itt} & 0.0036  & 0.0036 & 0.0012 & 0.85 & 0.940 & 0.0012 & 0.00004 \\
		\eqref{equation:ITTP} &  \eqref{equation:ittp} & 0.1616 & 0.1616 & 0.0500 & 0.87 & 0.938 & 0.0514 & 0.0016\\
		\hline
	\end{tabular}
	\caption{Comparison of subgroup and population-level estimators. The estimators from top to bottom are estimating equation and likelihood based estimation of the subgroup cancer mortality hazard ratio, absolute and proportional subgroup cancer mortality reduction, and absolute and proportional population cancer mortality reduction. SE stands for standard error, MCsd for Monte Carlo standard deviation and MCE for Monte Carlo error.}\label{table}
\end{table}

\subsection{Simulating from constant transition rates}\label{section:constant_rates}

To check the performance of the proposed methods under settings not resembling a typical randomized screening trial with large size and rare events, we also simulated scenarios with smaller sample size and relatively more common events. In particular, we simulated scenarios with more than $50\%$ in the screening arm early diagnosed through (compared to $2.4\%$ in NLST) and among them majority experiencing cancer-specific death. The screening arm state transitions were simulated from the following constant transition rates, with follow-up type I censored at $t=7$:
\[\lambda^{11}(t) = 
\begin{blockarray}{c@{\hspace{1pt}}rrrr@{\hspace{4pt}}}
	& 1   & 2   & 3 & 4 \\
	\begin{block}{r@{\hspace{1pt}}|@{\hspace{1pt}}
			|@{\hspace{1pt}}rrrr@{\hspace{1pt}}|@{\hspace{1pt}}|}
		1 & . & 0.2280 & 0.1148 & 0.0168\\
		2 & 0 & . & 0.1980   & 0.0111 \\
		3 & 0 & 0 & . & 0 \\
		4 & 0 &  0 & 0  & . \\
	\end{block}
\end{blockarray}\]
For the control arm, all the intensities in the matrix were the same, except for $ \lambda^{10}_{23}(t) = 0.1980 \, \times \theta$, where $\log \theta$ was fixed to $0.4700$ (i.e. $\theta = 1.6$). From the intensity matrices, we generated screening trial data using the algorithm of Section \ref{section:algorithm} with sample sizes $n = 500, 800,$ and $1000$ and estimated the hazard ratios in the screening detectable subgroup using the two methods described in Section \ref{section:methods}. We did not consider the risk reduction type measures here. The results for a scenario without unmeasured confounding are presented in Table \ref{table2} (rows with $\beta=0$).

\begin{table}[ht!]
	\centering
	\resizebox{1\linewidth}{!}{
		\begin{tabular}{l|lllllllll}
			\hline
			\multirow{2}{*}{$\beta=0$} &\multirow{2}{*}{Estimator}  & Causal & \multirow{2}{*}{Truth} & Point & \multirow{2}{*}{SE} & \multirow{2}{*}{Power} & \multirow{2}{*}{Coverage} & \multirow{2}{*}{MCsd} &\multirow{2}{*}{MCE} \\
			&& contrast & & estimate & & & & & \\
			\hline
			$n = 500$& \eqref{equation:cumulative_control_cancer} & $\log \theta$ & 0.4700 & 0.4500  & 0.2467 & 0.44 & 0.95 & 0.2363 & 0.0106 \\
			& \eqref{equation:lik} & $\log \theta$ & 0.4700 & 0.4544 & 0.2324 & 0.50 & 0.95 & 0.2216 & 0.0099\\
			\hline
			$n = 800$ & \eqref{equation:cumulative_control_cancer} & $\log \theta$ & 0.4700 & 0.4708  & 0.1931 & 0.69 & 0.96 & 0.1828 & 0.0082 \\
			& \eqref{equation:lik} & $\log \theta$ & 0.4700 & 0.4674 & 0.1807& 0.75 & 0.96 & 0.1760 & 0.0079\\
			\hline
			$n = 1000$ & \eqref{equation:cumulative_control_cancer} & $\log \theta$ & 0.4700 & 0.4730  & 0.1698 & 0.82 & 0.95 & 0.1652 & 0.0074 \\
			& \eqref{equation:lik} & $\log \theta$ & 0.4700 & 0.4686 & 0.1600 & 0.85 & 0.95 & 0.1601 & 0.0072\\
			\hline
			\multirow{2}{*}{$\beta=0.34$} &\multirow{2}{*}{Estimator}  & Causal & \multirow{2}{*}{Truth} & Point & \multirow{2}{*}{SE} & \multirow{2}{*}{Power} & \multirow{2}{*}{Coverage} & \multirow{2}{*}{MCsd} &\multirow{2}{*}{MCE} \\
			&& contrast & & estimate & & & & & \\
			\hline
			$n = 500$ & \eqref{equation:cumulative_control_cancer}  & $\log \theta$ & 0.3900 & 0.4034  & 0.2506 & 0.37 & 0.95 & 0.2256 & 0.0101\\ 
			& \eqref{equation:lik} & $\log \theta$ & 0.3900 & 0.4052 & 0.2275 & 0.43 & 0.96 & 0.2098 & 0.0094\\
			\hline
			$n = 800$ & \eqref{equation:cumulative_control_cancer}  & $\log \theta$ & 0.3941 & 0.3807  & 0.1827 & 0.57 & 0.96 & 0.1740 & 0.0078 \\ 
			& \eqref{equation:lik} & $\log \theta$ & 0.3941 & 0.3796 & 0.1681 & 0.62 & 0.97 & 0.1632 & 0.0073\\			
			\hline
			$n = 1000$ & \eqref{equation:cumulative_control_cancer}  & $\log \theta$ & 0.4000 & 0.4268  & 0.1658 & 0.77 & 0.95 & 0.1642 & 0.0073 \\ 
			& \eqref{equation:lik} & $\log \theta$ & 0.4000 & 0.4221 & 0.1535 & 0.82 & 0.95 & 0.1528 & 0.0068\\
			\hline		
			\multirow{2}{*}{$\beta=0.47$} &\multirow{2}{*}{Estimator}  & Causal & \multirow{2}{*}{Truth} & Point & \multirow{2}{*}{SE} & \multirow{2}{*}{Power} & \multirow{2}{*}{Coverage} & \multirow{2}{*}{MCsd} &\multirow{2}{*}{MCE} \\
			&& contrast & & estimate & & & & & \\
			\hline
			$n = 500$ & \eqref{equation:cumulative_control_cancer}  & $\log \theta$ & 0.3794 & 0.3766  & 0.2633 & 0.24 & 0.96 & 0.2199 & 0.0098 \\ 
			& \eqref{equation:lik} & $\log \theta$ & 0.3794 & 0.3744 & 0.2288 & 0.37 & 0.96 & 0.1968 & 0.0088\\
			\hline
			$n = 800$ & \eqref{equation:cumulative_control_cancer}  & $\log \theta$ & 0.3819 & 0.3436  & 0.1849 & 0.46 & 0.95 & 0.1697 & 0.0076 \\ 
			& \eqref{equation:lik} & $\log \theta$ & 0.3819 & 0.3423 & 0.1682 & 0.53 & 0.95 & 0.1604 & 0.0072\\			
			\hline
			$n = 1000$ & \eqref{equation:cumulative_control_cancer}  & $\log \theta$ & 0.3876 & 0.4044  & 0.1691 & 0.73 & 0.96 & 0.1554 & 0.0069 \\ 
			& \eqref{equation:lik} & $\log \theta$ & 0.3876 & 0.4014 & 0.1540 & 0.77 & 0.96 & 0.1464 & 0.0065\\
			\hline	
	\end{tabular}}
	\caption{Comparison of subgroup hazard ratio estimates with varied sample sizes, based on $500$ simulated datasets. The two estimators correspond to estimating equation and likelihood based estimation. SE stands for standard error, MCsd for Monte Carlo standard deviation and MCE for Monte Carlo error.\label{table2}}
\end{table}

Results in Table \ref{table2} indicate that the two estimators again perform very similarly. With $n=500$, they show some small sample bias, which disappears with increasing sample size. The bootstrap standard errors give a reasonable approximation of the sampling distribution standard deviation. These results indicate that when the modeling and identifying assumptions are satisfied, the methods perform reasonably also with smaller sample size and more common events.

To study how the methods perform in the presence of unmeasured confounding, we modified the above transition intensities for transitions $1 \rightarrow 2$, $1 \rightarrow 3$ and $2 \rightarrow 3$ as $\lambda_{12}^1(t) \exp (\beta U)$, $\lambda_{13}^1(t) \exp (\beta U)$ and $\lambda_{23}^{11}(t) \exp (\beta U)$, where $U \sim \textrm{Bernoulli}(0.5)$ is an unmeasured baseline covariate. This represents a scenario where some individuals are at higher risk for cancer diagnosis and cancer mortality, but under which the instrumental variable assumptions are still satisfied. However, because the data are simulated from a conditional generating mechanism, some degree of model misspecification is expected in the estimators. In addition to the proportionality of the marginal effect, the estimators use a Markov multi-state model for calculation of the state occupancy probabilities. While the Markov assumption can be relaxed, here we study its effect on the estimates. The true marginal effects were approximated as explained in Section \ref{section:algorithm}.

The results with $\beta=0.34$ and $\beta=0.47$ are also summarized in Table \ref{table2}. They show as expected that the true marginal effects are now smaller compared to the conditional effect used in the simulation. The points estimates were close to the true marginal effects under most of the scenarios and sample sizes, with coverage probabilities close to nominal. This suggests that the estimates are free of confounding bias, as is expected of instrumental variable estimators.

\section{Illustration using NLST data}\label{section:illustration}
The NLST data were briefly introduced in Section \ref{section:conventional}. The original analysis followed the ITS principle. Here we are interested in quantifying the effect of the early vs late treatments among the screening detectable subgroup in terms of the subgroup cancer-specific mortality hazard ratio. For this purpose, we carried out a secondary analysis of the NLST data using the estimating equation approach based on \eqref{equation:cumulative_control_cancer}. We estimated the hazard ratio as a function of the fixed time point $t$, to determine if this effect stabilizes or dilutes over time. The confidence intervals were generated using $500$ bootstrap replicates.

\begin{figure}[ht!]
	\centering
	\includegraphics[width=\textwidth]{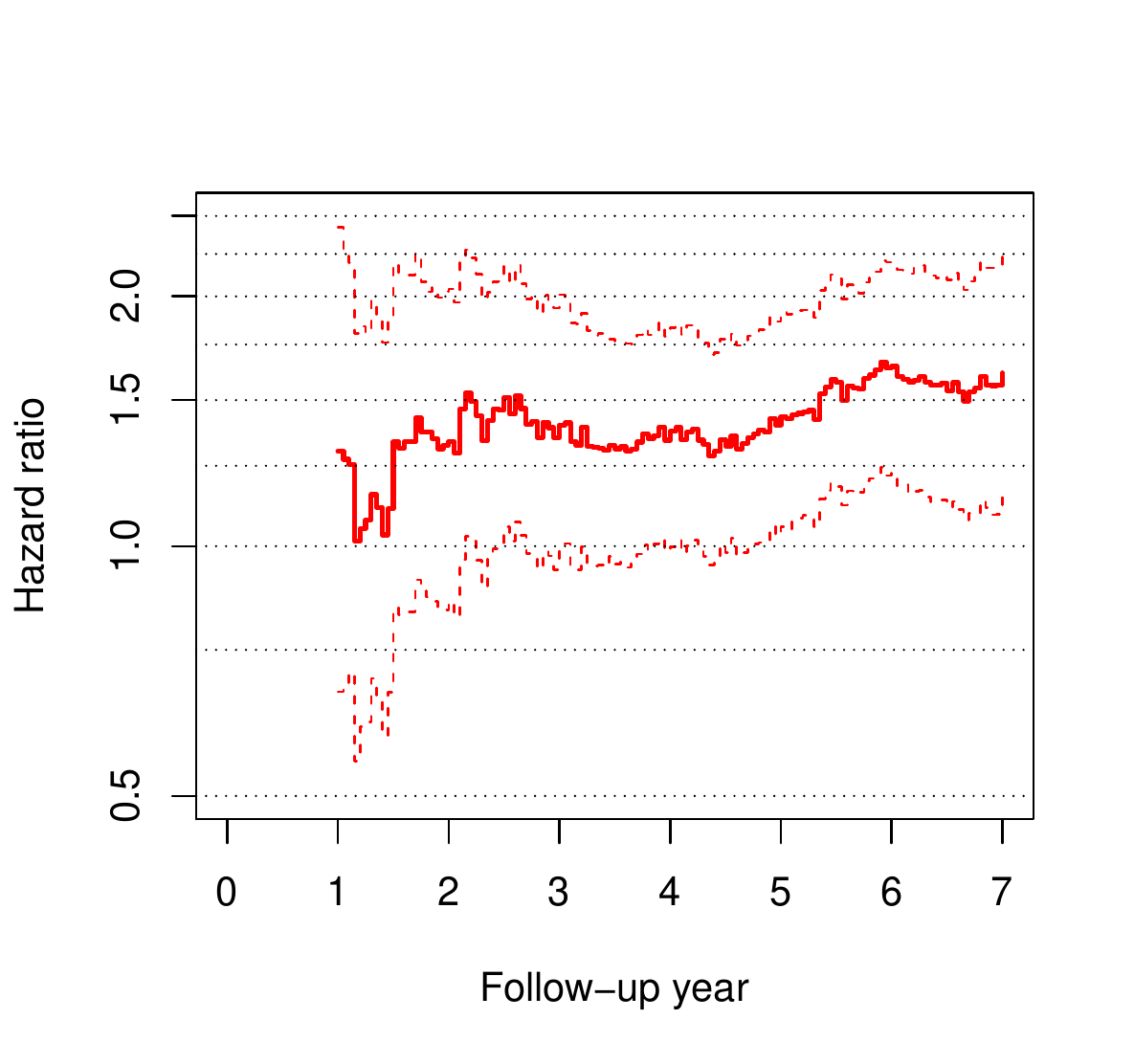}
	\caption{Estimated effect of delayed versus early treatment on lung cancer mortality among the screening-detectable subgroup. The dashed lines are 95\% bootstrap confidence bands. Estimates are generated as a function of time, using the follow-up data up to a given time point $t$. The estimates are noisy in the first year of follow-up and are shown from the second year.}
	\label{fig:HR}
\end{figure}

The results are presented in Figure \ref{fig:HR}. This demonstrates that the hazard ratio becomes statistically significant in the fifth year of follow-up, reaches a maximum of $1.668$ at around 6 years, and levels after that. Overall the effect estimate is relatively stable over time which may lend some support towards the proportionality of this effect. The hazard ratio estimate at 7 years indicates a $100\% \times (1-1/1.62) = 38.3\%$ cancer-specific mortality hazard reduction due to the early treatments. This can be contrasted to the reduction in the empirical cancer mortality rate between screening and control arms in the trial which after 7 years of follow-up was $1-(469/171,412)/(552/170,355)$, or around 15.6\%. The reason for the difference between these two measures is that the latter combines the effect of early treatments after early diagnosis with factors influencing early diagnosis itself, including the screening regimen, screening technology and non-compliance. The statistical significance of the two effects is similar based on the NLST data; even though the subgroup effect is larger in magnitude, it is evaluated in a smaller subpopulation, and these two aspects counterbalance each other compared to the smaller ITS effect in the entire study population.

In addition to reporting the estimate as a function of time and at the end of the follow-up, for choosing a single number effect estimate to report, we also applied the two approaches suggested in Section \ref{section:equation} for choosing the timepoint for estimation, namely the time minimizing the empirical variance of the estimate and the inverse variance weighted average. For these, we considered a grid of timepoints from 1 to 7 years with $0.05$ year intervals. We found that the minimum empirical variance is achieved at $t=4.55$ with hazard ratio of $1.36$, while the inverse variance weighted average hazard ratio was $1.43$, suggesting that the choice of the timepoint made relatively small difference to the estimate.

\section{Discussion}\label{section:discussion}

In this paper, we proposed a simplified multi-state model to characterize screening trials, and proposed a new measure (i.e. structural cause-specific hazard ratio) to quantify the impact of early treatments compared to delayed treatments in the screening-detectable subgroup. We note that the subgroup is specific to the screening regimen and screening technology implemented in the trial that is used for the estimation, as the effect is formulated in terms of a hypothetical intervention trial that shares the same regimen and technology. Transferability of the subgroup effect to different settings with different screening technology or regimen, disease incidence and non-adherence rate requires additional assumptions. Nevertheless, the subgroup effect is arguably more transferable compared to the ITS effect, which directly depends on all of these factors. We presented two alternative estimators for the subgroup effect. In addition to comparing properties of estimators, the simulation model may have use in planning new trials targeted for the subgroup effects. While the estimation methods we proposed allow for flexible estimation of the quantities not of interest, the effect of interest is quantified in terms of a single parameter. As a limitation of the present work, we did not propose methods for estimating time-dependent effects, but outline some ideas below.

While we contend that the cause-specific hazard ratio can serve as a useful summary measure of the magnitude of the effect, and simplifies accommodating competing risks into the analysis,
we do note that the use of marginal hazard ratios as a causal parameters has been criticized by several authors \citep[e.g.][]{aalen2015}, who point out even in randomized trials where the arms are initially balanced with respect to observed or unobserved baseline characteristics, this balance is lost over the course of the follow-up as more high risk individuals get removed from the risksets. However, as outlined by \citet{young2020causal}, counterfactual cause-specific hazards can always be formulated in terms of the underlying potential event time/event type outcomes in the two arms of a trial. Because these quantities are defined conditional on the past, it should be noted that the corresponding hazard ratio being one at a given time does not necessarily imply the absence of a causal effect, but rather some combination of causal effect and selection. If a strictly causal interpretation is sought, one could use the subgroup mortality risk reduction as the measure instead. 

With these reservations, we suggest interpreting the cause-specific hazard ratio as a summary measure of the early treatment effect over time. However, we note that the proportionality is more plausible for the subgroup early treatment effect since this effect begins from the early diagnosis, compared to the use of cancer mortality rate ratio to compare the screening and control arms, which is dependent on the screening regimen, among other things. In particular, when the screening is discontinued, the screening effect eventually disappears in the full cohort and continuing the follow-up will dilute the proportional effect. The subgroup effect measure does not suffer from this limitation, as no more subgroup members are accumulated after the screening discontinues. In fact, we suggest that the present proposal can resolve some of the controversies in characterizing screening effect through a hazard/rate ratio, discussed for example by \citet{liu2013,liu2015conditional,hanley2018disaggregating,habbema2018statistical}. Rather than the screening effect in the entire population, what could be plausibly assumed proportional is the early treatment effect in the subgroup. On the other hand, the usual residual type diagnostics for proportionality violations are not applicable under the instrumental variable type estimation methods; diagnosing violations of proportionality under the current framework is a topic for further research. As one possible approach, the likelihood-based method could be extended to make use of the observed event times in the control arm, making it potentially informative of the time-varying effects. For this purpose, the likelihood contributions in the control arm could be evaluated at the minimum of observed time of death and censoring time $T_i$ as
\begin{align}\label{equation:liknew}
	L(\theta) &\stackrel{\theta}{\propto} \prod_{\{i:Z_i = 0\}} E[\textrm d N_{13}^0(T_i)]^{\textrm d N_{13}^{0}(T_i)}  E[\textrm d N_{14}^0(T_i)]^{\textrm d N_{14}^{0}(T_i)} \nonumber \\
	&\qquad\times \left(1-E[N_{13}^0(T_i)]-E[N_{14}^0(T_i)]\right)^{1-\textrm d N_{13}^0(T_i)-\textrm d N_{14}^0(T_i)},
\end{align}
where we can characterize for example the sub-density of the cancer-specific death time through
\begin{align*}
	E[\textrm d  N_{13}^0(t)] &=  \exp \left\{-(\Lambda_{13}^0(t)+\Lambda_{14}^0(t))\right\} \lambda_{13}^0(t) \,\textrm dt \nonumber \\
	&= E[\textrm d  N_{13}^{0}(t) \mid N_{12}^1(t)=1] P(N_{12}^1(t)=1) \nonumber \\
	&\quad+ E[\textrm d  N_{13}^0(t) \mid N_{12}^1(t)=0]P(N_{12}^1(t)=0) \nonumber \\
	&\stackrel{\textrm{(i),(ii)}}{=} E[\textrm d  N_{23}^{10}(t) \mid N_{12}^1(t)=1] P(N_{12}^1(t)=1) \nonumber \\
	&\quad+  E[\textrm d  N_{13}^1(t) \mid N_{12}^1(t)=0] P(N_{12}^1(t)=0) \nonumber \\
	&= \int_{0}^{t} \exp \left\{-(\Lambda_{12}^{1}(u)+\Lambda_{13}^{1}(u)+\Lambda_{14}^1(u)) \right\} \lambda_{12}^{1}(u) \nonumber\\
	&\quad\times \exp \left\{- \left(\int_{u}^{t} [\lambda_{23}^{10}(v) + \lambda_{24}^{10}(v)] \,\textrm dv\right) \right\} \lambda_{23}^{10}(t) \,\textrm dt \,\textrm du \nonumber\\
	&\quad+ \exp \left\{-(\Lambda_{12}^1(t)+\Lambda_{13}^1(t)+\Lambda_{14}^1(t))\right\} \lambda_{13}^1(t) \,\textrm dt \\
\end{align*}
and further, under proportionality, 
\begin{align*}
	E[\textrm d N_{13}^0(t)]
	&\stackrel{\textrm{(iii)}}{=} \int_{0}^{t} \exp \left\{-(\Lambda_{12}^{1}(u)+\Lambda_{13}^{1}(u)+\Lambda_{14}^1(u)) \right\} \lambda_{12}^{1}(u) \nonumber\\
	&\quad\times \exp \left\{- \left(\int_{u}^{t} [\theta\lambda_{23}^{1}(v) + \lambda_{24}^{1}(v)] \,\textrm dv\right) \right\} \theta \lambda_{23}^{1}(t) \,\textrm dt\,\textrm du \nonumber\\
	&\quad+ \exp \left\{-(\Lambda_{12}^1(t)+\Lambda_{13}^1(t)+\Lambda_{14}^1(t))\right\} \lambda_{13}^1(t) \,\textrm dt,
\end{align*}
with the cumulative incidences in \eqref{equation:liknew} as before. This type of likelihood would allow testing the significance of interaction terms with time entered into the model. The likelihood could also be used to assess on which time scale proportionality fits the data better. It could also accommodate independent censoring which can be seen by introducing a further counting process characterizing the censoring events, and factoring out the corresponding terms from the likelihood. The likelihood could also be written for covariate conditional quantities, accommodating covariate dependent censoring. We note that because the event times in the screening and control arms are distinct, substituting estimators for the sub-densities from the screening arm of the trial would require smooth hazard estimates. This is in contrast to the methods we focused on earlier, where non-parametric estimators could be substituted in for the cumulative incidences.

The proposed estimating equation based method could accommodate independent censoring. It can be extended straightforwardly to the case where the censoring is conditional on baseline covariates, for example, standardizing over the empirical covariate distribution in the control arm of the trial, equation \eqref{equation:cumulative_control_cancer} can be expressed as
\begin{align*}
	\MoveEqLeft \frac{1}{n} \sum_{\{i: Z_i = 0\}} E[N_{13}^0(t) \mid x_i] \\
	&= \frac{1}{n}\sum_{\{i: Z_i = 0\}} \Bigg[ \int_{0}^{t} \exp \left\{-(\Lambda_{12}^{1}(u \mid x_i)+\Lambda_{13}^{1}(u \mid x_i)+\Lambda_{14}^1(u \mid x_i)) \right\} \lambda_{12}^{1}(u \mid x_i) \nonumber\\
	&\quad\times \int_{u}^{t} \exp \left\{- \left( \int_{u}^{v} [\theta \lambda_{23}^{1}(s\mid x_i) + \lambda_{24}^{1}(s \mid x_i)] \,\textrm ds\right) \right\} \theta \lambda_{23}^{1}(v \mid x_i) \,\textrm dv\,\textrm du \nonumber\\
	&\quad+ \int_{0}^{t} \exp \left\{-(\Lambda_{12}^1(u \mid x_i)+\Lambda_{13}^1(u \mid x_i)+\Lambda_{14}^1(u \mid x_i))\right\} \lambda_{13}^1(u \mid x_i) \,\textrm du \Bigg],
\end{align*}
where the covariate conditional cause-specific hazards can be estimated using Cox models, and the cumulative incidences on the left hand side either with Cox or Fine \& Gray models. In addition to allowing for covariate-dependent censoring, conditioning on the covariates may be useful for relaxing the instrumental variable assumptions outside of randomized trials, or improving the efficiency of the instrumental variable type estimators \citep{burgess2017review}. In this case the estimated hazard ratio will be covariate conditional. Alternatively, to preserve the original marginal estimand, the original estimating equation \eqref{equation:cumulative_control_cancer} could be used when combined with inverse probability of censoring weighted estimators for the other unknown quantities \citep[e.g.][]{howe2016selection}. We are currently pursuing these extensions.

\subsection*{Acknowledgment}

This work was supported by the Ontario Institute for Cancer Research through funding provided by the Government of Ontario (to SS) and by a Discovery Grant from the Natural Sciences and Engineering Research Council of Canada (to OS). The authors thank the National Cancer Institute (NCI) for access to NCI’s data collected by the National Lung Screening Trial. The statements contained herein are solely those of the authors and do not represent or imply concurrence or endorsement by the NCI.

\end{document}